\documentclass{aipproc}  \layoutstyle{6x9} \usepackage{ifthen}

\newboolean{usepdf} \newif\ifpdf \ifx\pdfoutput\undefined \pdffalse \else \pdfoutput=1 \pdftrue \fi

\ifpdf \setboolean{usepdf}{true} \else \setboolean{usepdf}{false} \fi

\graphicspath {{../documents/thesis/figures/}{./}}

\newcommand{\yr}{\, {\rm yr}}

\newcommand{\Msun}{\,\mathrm{M}_{\odot}}
\newcommand{\Lsun}{\,\mathrm{L}_{\odot}}

\newcommand{\halpha}{{\mathrm{H}\alpha}}

\newcommand{\mcrx}{\textit{Sunrise}}

 \usepackage{url} \usepackage{comment} %\usepackage{natbib}
\usepackage{astrojournals}

\begin {document} \title{Simulations of Dust in Interacting Galaxies} \author{Patrik Jonsson} {address={University of California, Santa Cruz}} \author{T. J. Cox} {address={Harvard-Smithsonian Center for Astrophysics}} \author{Joel R. Primack} {address={University of California, Santa Cruz}} \begin {abstract}
%NLX% end exclude from vocabulary builder  
  A new Monte-Carlo radiative-transfer code, \mcrx , is used to study  
the effects of dust in N-body/hydrodynamic simulations of   interacting
galaxies.  Dust has a profound effect on the appearance   of the
simulated galaxies.  At peak luminosities, $\sim 90 \%$ of the
bolometric luminosity is absorbed, and the dust   obscuration scales
with luminosity in such a way that the brightness   at UV/visual
wavelengths remains roughly constant.  A general   relationship between
the fraction of energy absorbed and the ratio   of bolometric
luminosity to baryonic mass is found.

     Comparing to observations, the simulations are found to follow a  
relation similar to the observed IRX-$\beta$ relation found by  
\citet{mhc99} when similar luminosity objects are   considered.  The
highest-luminosity simulated galaxies depart from   this relation and
occupy the region where local (U)LIRGs are found.   This agreement is
contingent on the presence of Milky-Way-like dust,   while SMC-like
dust results in far too red a UV continuum slope to   match
observations.

     The simulations are used to study the performance of
star-formation   indicators in the presence of dust.  The far-infrared
luminosity is   found to be reliable.  In contrast, the $\halpha$ and  
far-UV luminosity suffer severely from dust attenuation,   and dust
corrections can only partially remedy the situation.
%NLX% exclude from vocabulary builder
\end {abstract} \maketitle \section {Introduction}
%NLX% end exclude from vocabulary builder

Galaxy mergers are an important ingredient in the hierarchical picture
of galaxy formation.  They transform disks to spheroids, and may have
been responsible for forming a majority of the stars in the Universe
\citep{SPF}.  Interacting galaxies are also triggering the most
luminous starbursts in the Universe, in the Luminous and Ultraluminous
Infrared Galaxies. Numerical simulations have been used to study
interacting galaxies and the resulting starbursts \citep{MH96, Sp00},
but these simulations alone cannot predict what these objects would
look like to observers, as the spectacular bursts of star formation in
(U)LIRGS are completely hidden by interstellar dust. Calculating the
effect of dust in galaxies is complicated.  The mixed geometry of stars
and dust makes the dust effects geometry-dependent and nontrivial to
deduce. Because of this, a full radiative-transfer model is necessary
to calculate these effects realistically.  Previous studies of dust in
galaxies have, with few exceptions \citep{bekkishioya00a,
bekkishioya00b}, not used information from hydrodynamic simulations. 
This study applies a new Monte-Carlo radiative-transfer code, \mcrx ,
to the outputs from a comprehensive suite of N-body/hydrodynamic
simulations of merging galaxies \citep{tjthesisurl}.

  The N-body simulations of merging galaxies consist of a comprehensive
suite of hydrodynamic simulations using the GADGET N-body/SPH code. The
simulations study the effects of merger mass ratio, encounter orbit,
and progenitor galaxy structure, as well as different star-formation
and feedback prescriptions, on the ensuing starburst and the structure
of the merger remnant \citep{tjthesisurl}.  The merging galaxies have
been modeled to closely resemble observed spiral galaxies in the local
Universe. Examples of results from this study include the discovery of
a shock-driven gas outflow from some merging systems \citep{Cox04} and
an improved measure of how the intensity of the starburst scales with
merger mass ratio in minor mergers. Our simulations also show how a
disk reforms after the merger event, as in simulations by
\citet{springelhernquist04disk}.

For mergers with mass ratios smaller than 1:5, there is little induced
star formation.  Also, major mergers of smaller progenitor galaxies
have fundamentally different star-formation properties than larger
ones, with larger and more prolonged enhancements of star formation due
to the merger event.  Our suite of merging galaxy simulations is by far
the largest performed so far.

\section{Radiative-Transfer Model}

In order to calculate the effects of dust in these simulations, a new
Monte-Carlo radiative-transfer code, \mcrx , was developed
\citep{pjthesis}.  The geometry of stars and gas in the N-body
simulations and the star-formation history of the system are used as
inputs to the radiative-transfer calculation. To make calculations with
such complicated geometries feasible, \mcrx \ features an adaptive-mesh
refinement grid and is shared-memory parallel.  For the stellar
emission, SEDs from the Starburst99 population synthesis model
\citep{leithereretal99} are used.  The dust model is taken from
\citet{weingartnerdraine01}, and dust is assumed to trace metals in the
simulations.  Currently, the infrared dust emission is not calculated
self-consistently.  Instead, the infrared templates of
\citet{devriendtetal99} are used.  A self-consistent dust emission
calculation is planned for the future.  In addition to using N-body
simulations as inputs, the code can be used to solve problems specified
in other ways.  As a service to the community, \mcrx \ is being
released to the public under the GNU General Public
%NLX% exclude from vocabulary builder
License\footnote{The \mcrx\ web site is   \url{http://sunrise.familjenjonsson.org}.}. 
%NLX% end exclude from vocabulary builder

\section{Simulation Results}

\begin{figure}   \includegraphics[width=0.32\textwidth]{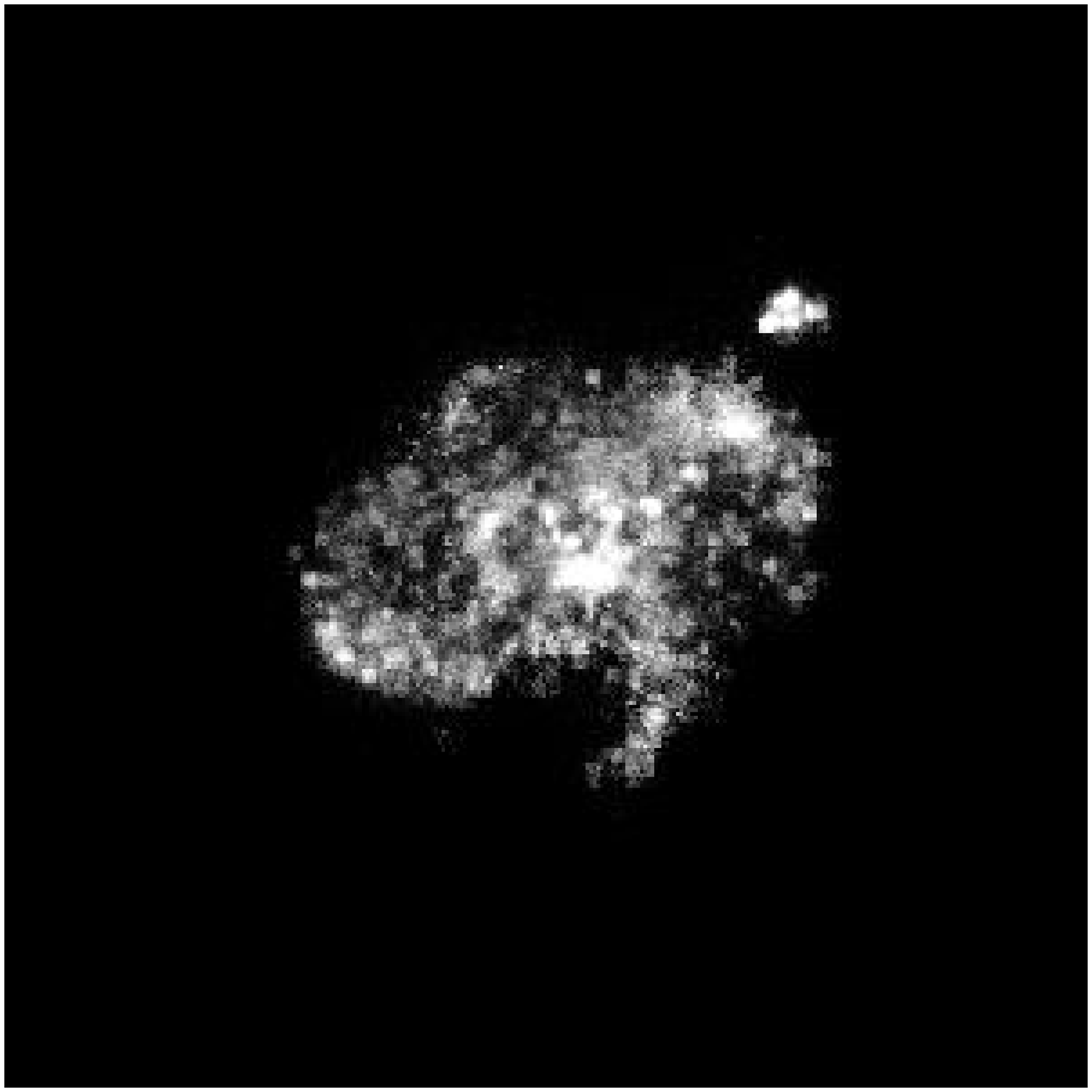}   \includegraphics[width=0.32\textwidth]{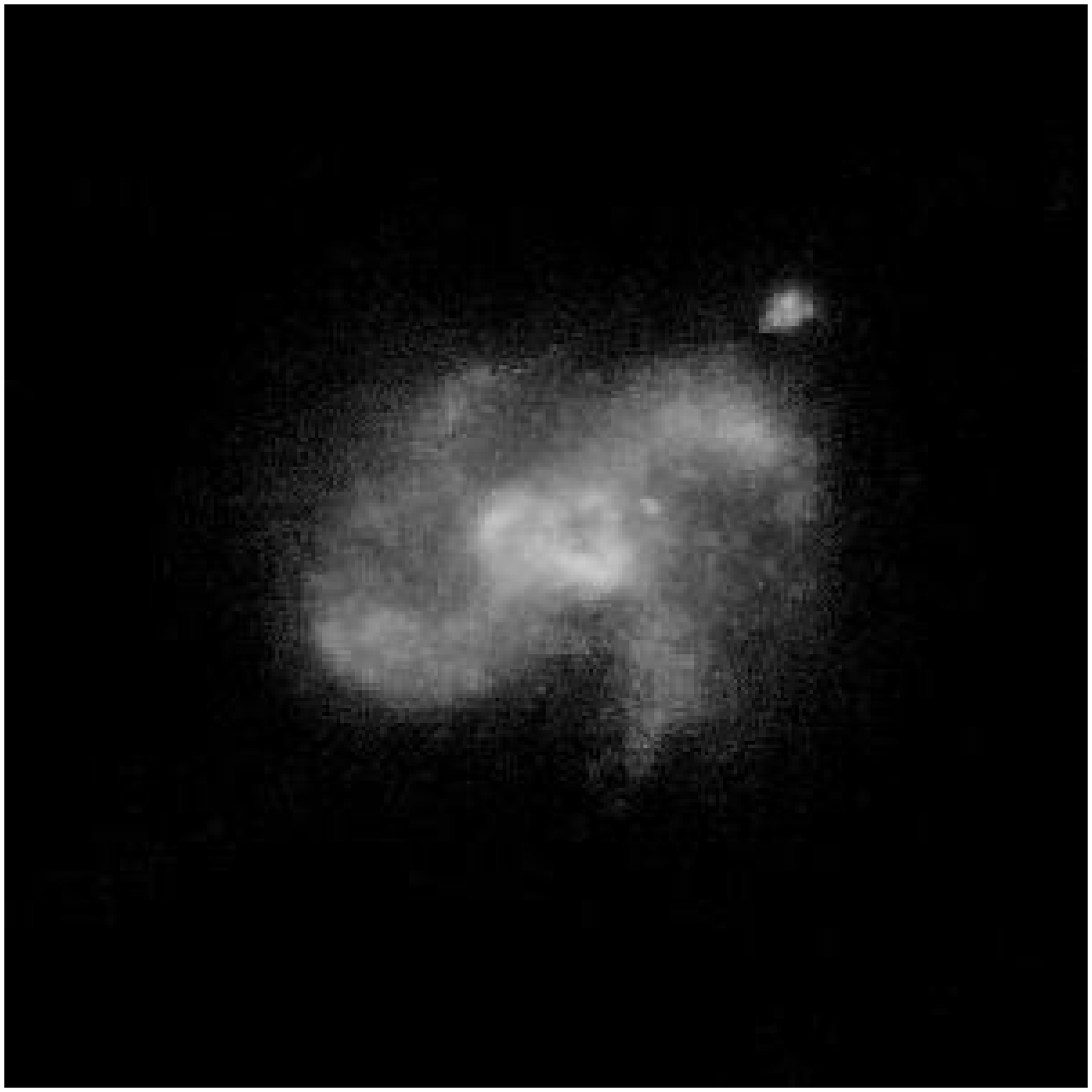}   \includegraphics[width=0.32\textwidth]{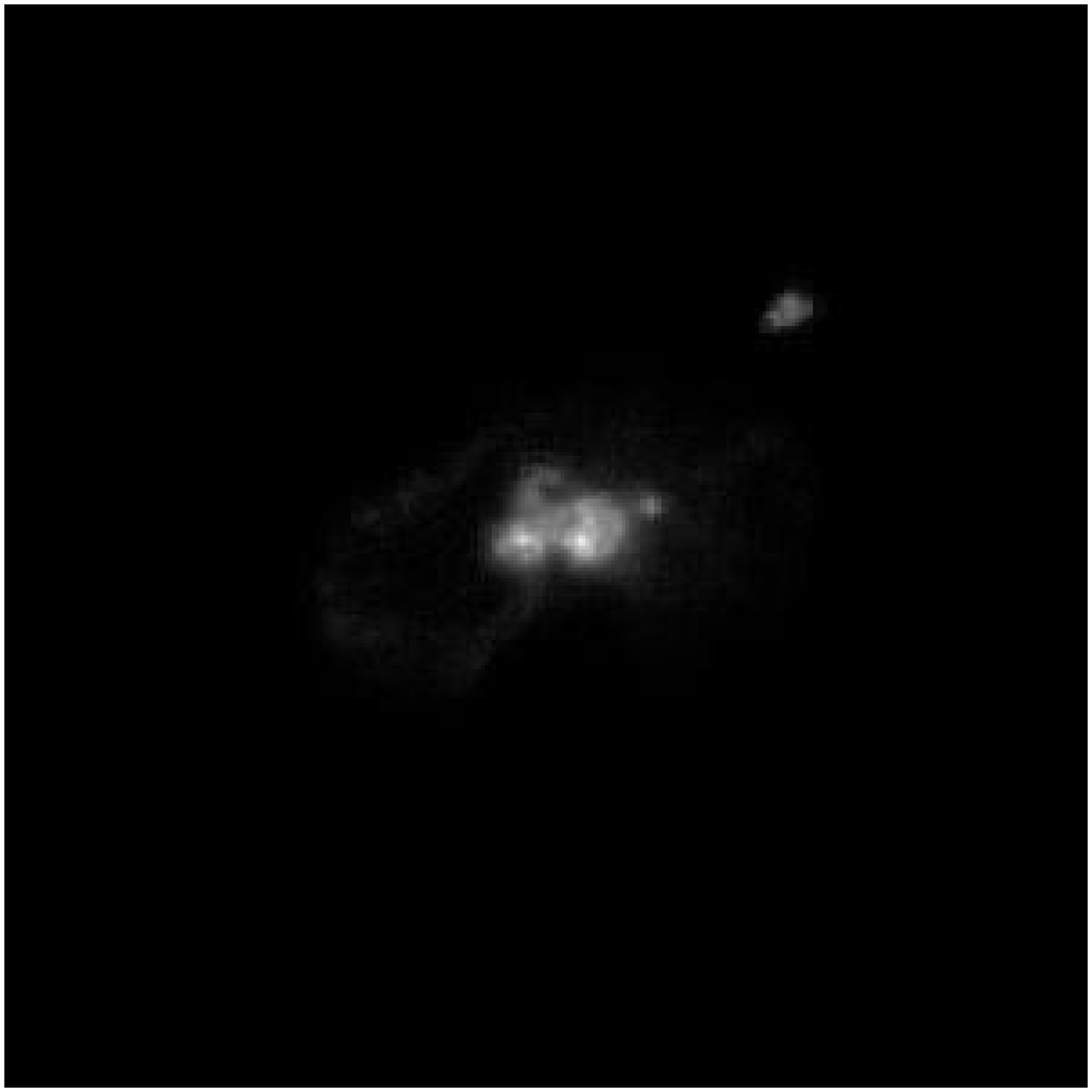}   \caption{     \label{multiband_images}     Images of a simulated late-stage galaxy merger showing, from left to     right, GALEX FUV band, SDSS \textit{r} band and infrared dust emission.     Because the radiative-transfer model currently does not calculate     a spatially dependent infrared SED, the IR image corresponds to     bolometric dust luminosity. The images cover 100\,kpc. While the     UV/visual images show a peculiar, obviously dusty, galaxy, the IR     image shows that the cores of the progenitor galaxies are still     distinct and vigorously star forming.  Images like these can be     compared to observations of interacting systems.  } \end{figure}

Radiative-transfer calculations of over 20 simulated disk-galaxy major
mergers have been completed, resulting in over 11,000 simulated images
and spectra \citep{pjthesis}. Example images are shown in
Figure~\ref{multiband_images}, and movies of entire merger simulations
are available on the
%NLX% exclude from vocabulary builder
Internet\footnote{\url{http://sunrise.familjenjonsson.org/thesis}}.
%NLX% end exclude from vocabulary builder
A smaller number of simulations of the isolated progenitor galaxies
have also been done.

%NLX% exclude from vocabulary builder
%NLX% end exclude from vocabulary builder

  The radiative-transfer calculations show that dust has a profound
effect on the appearance of the simulated systems.  At the most
luminous phase of the merger, $\sim 90 \%$ of the bolometric luminosity
is absorbed by dust.  The dust attenuation scales with luminosity in
such a way that the brightness at UV/visual wavelengths remains roughly
constant throughout the merger event, even though the bolometric
luminosity of the system increases by a factor of 4 due to the
merger-driven starburst.  A general relationship between the fraction
of energy absorbed and the ratio of bolometric luminosity to baryonic
mass is found to hold in galaxies with metallicities $> 0.7 Z_\odot$
over a factor of 100 in mass.

%NLX% exclude from vocabulary builder
\begin{figure}[t] \resizebox{\columnwidth}{!}{     \includegraphics[width= 0.4\textwidth]{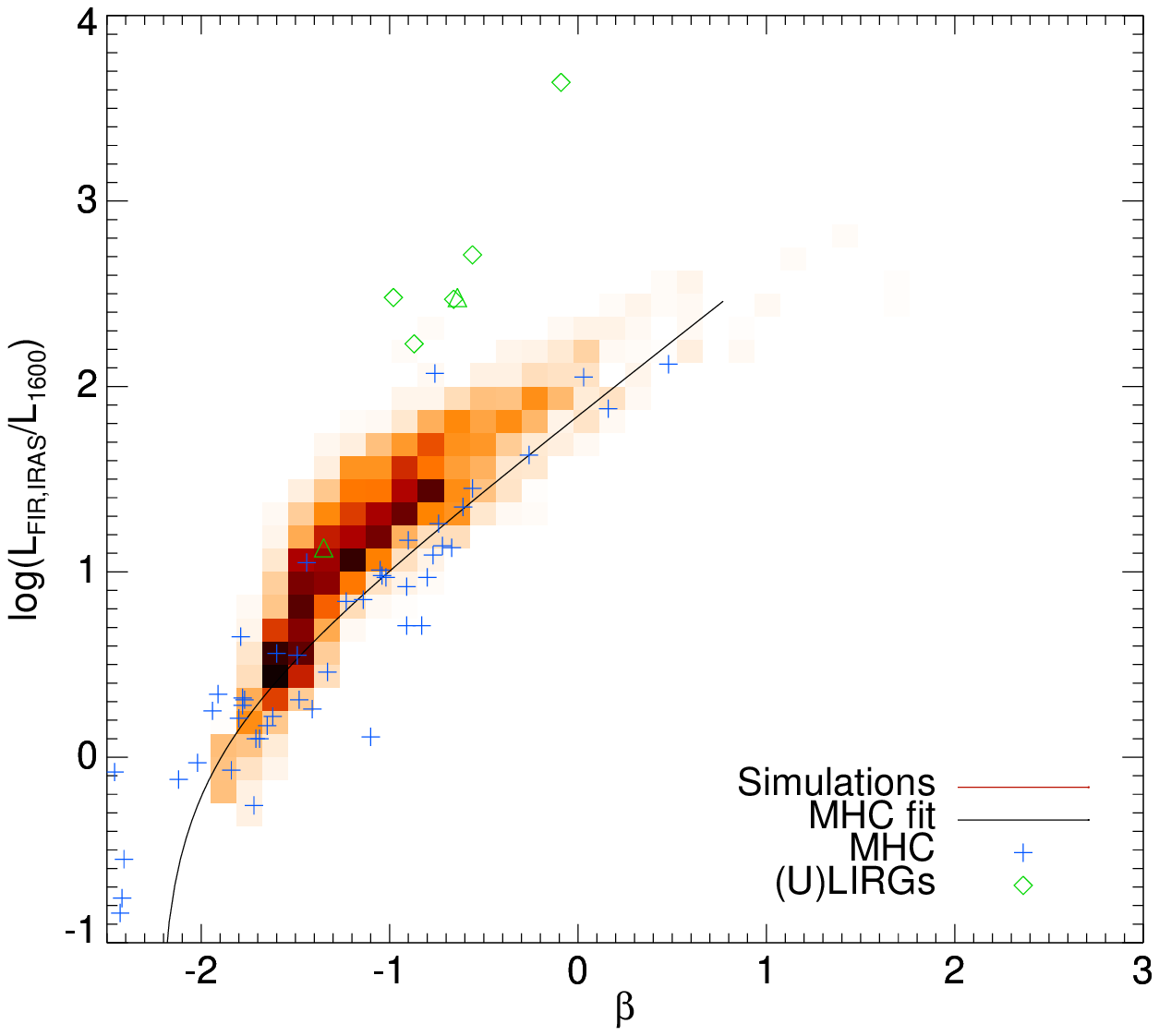}     \includegraphics[width= 0.4\textwidth]{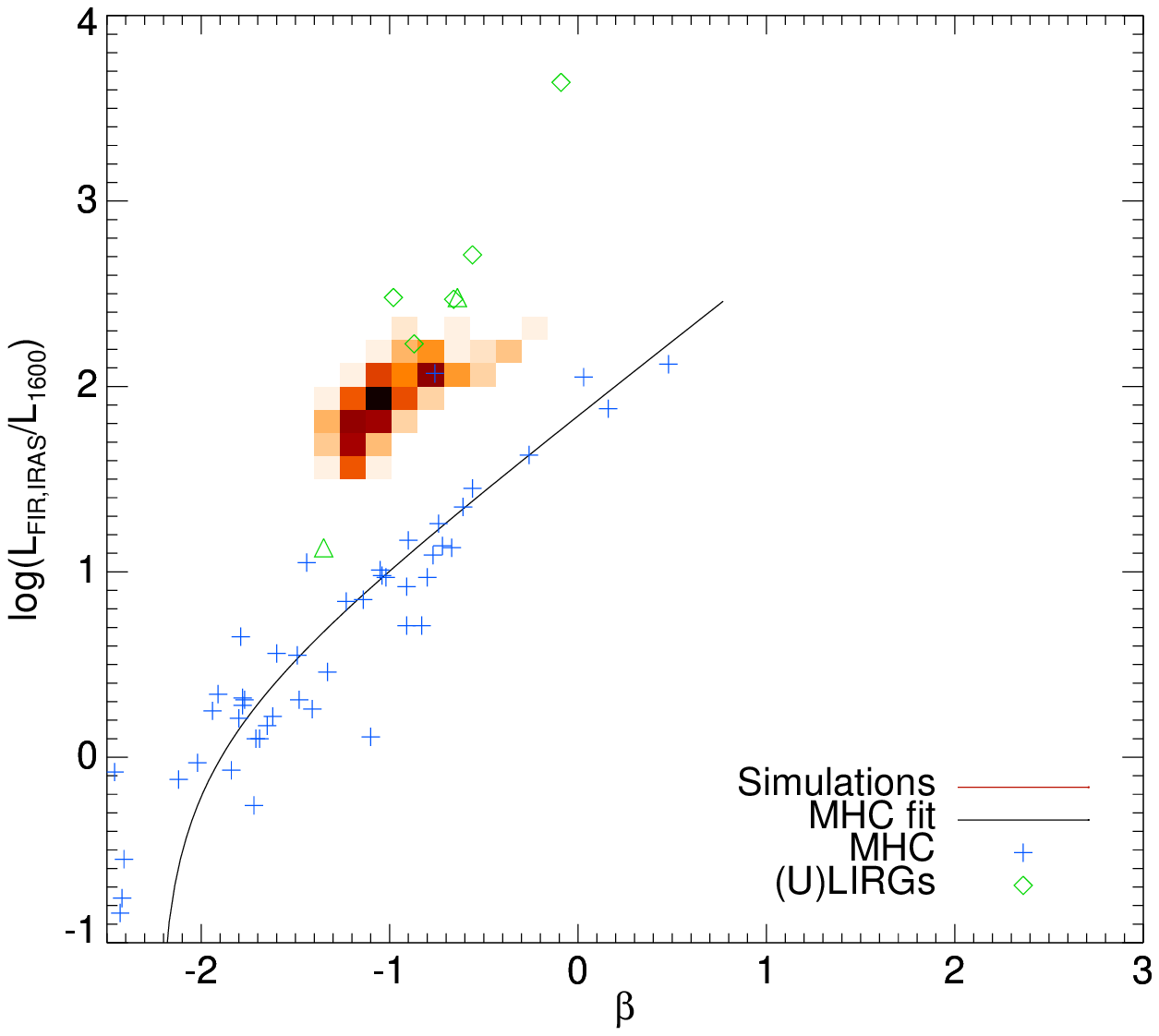} }   \caption{     The IRX-$\beta$ relation of the simulations (shaded region),     compared to the results from \citet*[][crosses]{mhc99} and     \citet[][diamonds/triangles]{goldaderetal02}.  On the left, only     simulated galaxies with bolometric luminosity $L_{\mathrm{bol}} <     2\cdot 10^{11}\Lsun$ have been included. This low-luminosity     sample agrees fairly well with the MHC correlation.  On the right,     only the highest-luminosity simulated galaxies, with     $L_{\mathrm{bol}} > 7\cdot 10^{11}\Lsun$ have been included.     These points depart completely from the MHC galaxies and instead     occupy the region of (U)LIRGs from the \citet{goldaderetal02}     sample.  }     \label{plot_irx_beta_luminosity} \end{figure}
%NLX% end exclude from vocabulary builder

The accuracy with which the simulations describe observed starburst
galaxies is evaluated by comparing them to observations by
\citet{mhc99} and \citet{h98}.  The simulations are found to follow a
relation similar to the IRX-$\beta$ relation found by \citet{mhc99}
when similar luminosity objects are considered. The highest-luminosity
simulated galaxies depart from this relation and occupy the region
where local (U)LIRGs are found.  These results are shown in
Figure~\ref{plot_irx_beta_luminosity}.  Comparing to the \citet{h98}
sample, the simulations are found to obey the same relations between UV
luminosity, UV color, IR luminosity, absolute blue magnitude and
metallicity as the observations.  This agreement is contingent on the
presence of a realistic mass-metallicity relation, and Milky-Way-like
dust.  In contrast with earlier studies \citep{gcw97}, SMC-like dust
results in far too red a UV continuum slope to match observations.  On
the whole, the agreement between the simulated and observed galaxies is
impressive considering that the simulations have not been fit to agree
with the observations, and we conclude that the simulations provide a
realistic replication of the real universe.

The simulations are then used to study the performance of
star-formation indicators in the presence of dust.  The far-infrared
luminosity is found to be a reliable tracer of star formation, as long
as the star-formation rate is larger than about $1 \Msun / \rm { \yr
}$.  In contrast, the $\halpha$ and far-ultraviolet luminosities suffer
severely from dust attenuation, as expected. Published dust corrections
based on the Balmer line ratios \citep{calzettietal94} or the
ultraviolet spectral slope \citep{bellkennicutt01} only partially
remedy the situation, still underestimating the star-formation rate by
up to an order of magnitude.

\section{acknowledgments}

PJ has been supported by a University Collaborative Grant through the
Institute for Geophysics and Planetary Physics (IGPP).  This work used
resources of the National Energy Research Scientific Computing Center
(NERSC), which is supported by the Office of Science of the U.S.
Department of Energy.

%NLX% exclude from vocabulary builder
\bibliographystyle{aipproc} \bibliography{../documents/thesis/thesis,../documents/thesis/tj,../documents/cv/bib,this} \end {document}